\documentstyle[12pt]{article}
\let\section=\subsection     \let\subsection=\subsubsection                %%
       
\begin{document}
\input epsf
\begin{center}
%\eqsec  % uncomment this line to get equations numbered by (sec.num)
{\large \bf HARD PHOTONS BEYOND PROTON-NEUTRON BREMSSTRAHLUNG IN HEAVY-ION
COLLISIONS} \footnote{Invited talk given at the NATO Advanced Research 
Workshop on the Structure of Mesons, Baryons and Nuclei, 
Krakow (Poland), May 26 - 30, 1998}\\[2mm]
K.~GUDIMA and M.~P{\L}OSZAJCZAK \\[5mm]
{\small \it Grand Acc\'{e}l\'{e}rateur National d'Ions Lourds (GANIL)\\ 
CEA/DSM -- CNRS/IN2P3, B.P. 5027, F-14076 Caen Cedex 05, France \\[8mm]}
\end{center}

\begin{abstract}\noindent
We report on the study of extremely high energy 
photons, pions and etas, produced in intermediate energy 
heavy-ion collisions. Possibility of imaging the final-state phase space in
these collisions by the Bose-Einstein correlations for photons is
critically examined.
\end{abstract}
%\PACS{25.70.-z,13.75.Cs}
 
\vfill
\newpage
 
\section{Introduction}
Measurement in heavy-ion (HI) 
collisions of hard photons of energies several times
higher than the beam energy per nucleon, has stimulated large interest in the
development of an appropriate theoretical framework for the 
description of rare energy fluctuations 
which could account for the measured properties of particles. In this context,
the success of the stochastic multiplicative
processes\cite{guardamar} in describing the subthreshold particle yields came
as a surprise. 
Nucleon in the transient state of the HI collision is
described by the stochastic equation :
$dp/dt = F + G(p) \lambda $~,
where $\lambda $~ is a random Gaussian process ($<\lambda > = 0, 
<{\lambda }^2> =
{\sigma}^2 $) accounting for fluctuating nuclear environment, and
$G(p) = g_0 + g_1p + \cdots $~~ is a smooth function of the particle momentum.
In the leading order, 
the time evolution of the relative momentum of two particles is :
$d({\delta p})/dt \sim \alpha (\delta p)\lambda $~
and, hence,  the relative kinetic energy evolves as :
\begin{eqnarray}
\label{eq1}
\frac{d}{dt}E_{NN} \sim 2\alpha E_{NN}\lambda ~~~ \ .
\end{eqnarray} 
The relative energy of the nucleon-nucleon pair in Eq. (\ref{eq1}) 
is a stochastic,
short-correlation quantity driven by the multiplicative noise\cite{guardamar,
zeldowicz}. Taking
$\varepsilon (t=0) \equiv \delta (E_{NN} - E_{CM}/A)$ as the initial condition,
\begin{figure}[tbh]
\begin{center}
\leavevmode
\epsfxsize=14.3cm
%\epsfysize=9.0cm
%\epsfverbosetrue
\epsfbox{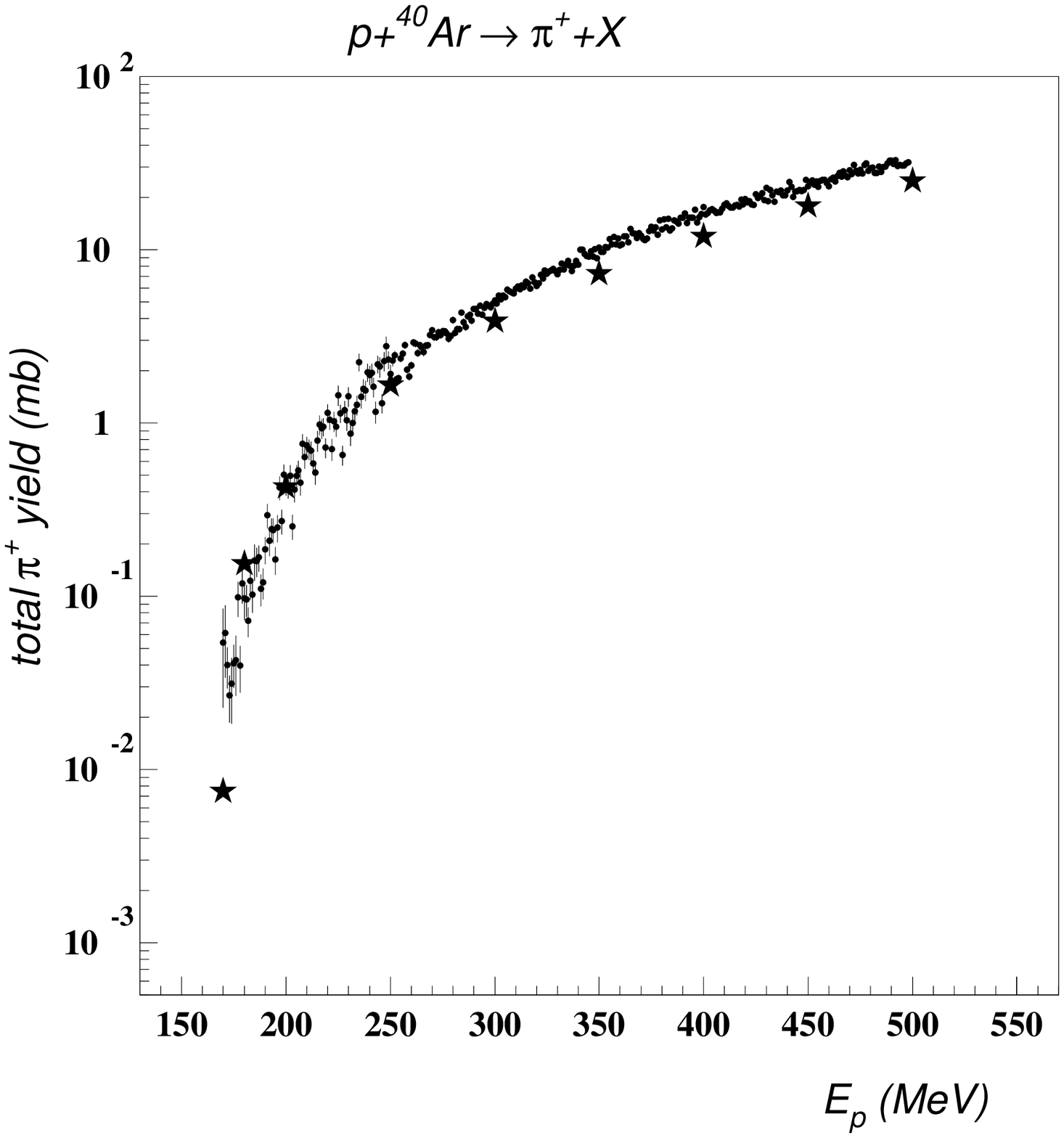}
%\epsffile{plo_fig1.eps}
\end{center}
{\begin{small} 
Fig.~1. Experimental energy and angle integrated yield of 
${\pi }^{+}$ from  $p + Ar$ reactions
is compared with the results of DCM
(stars). 
\end{small}}
\label{fig1}
\end{figure}
%%%%
\begin{figure}[tbh]
\begin{center}
\leavevmode
\epsfxsize=14.3cm
%\epsfysize=9.0cm
%\epsfverbosetrue
\epsfbox{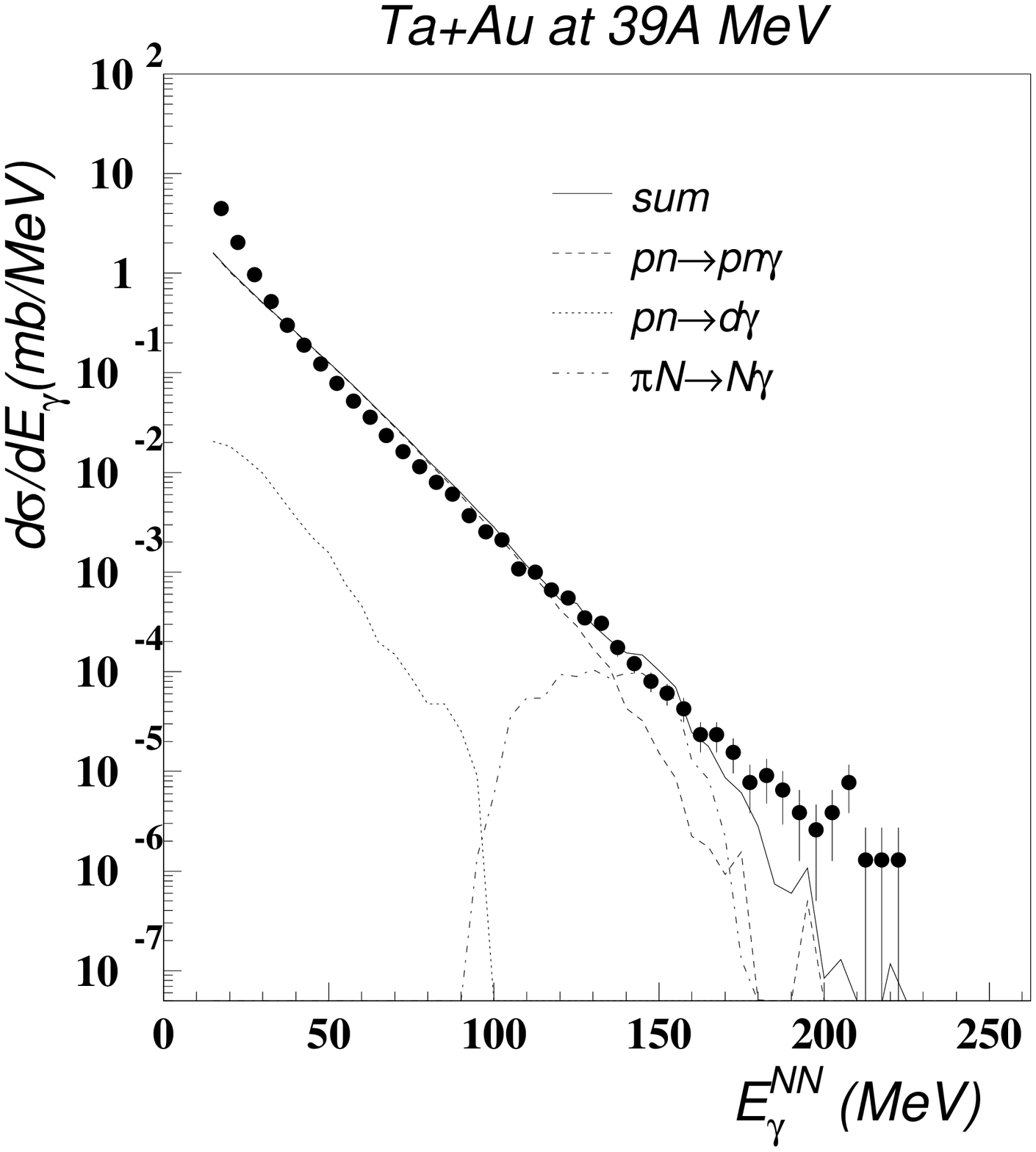}
%\epsffile{plo_fig2.eps}
\end{center}
%\vspace*{-1cm}
{\begin{small} 
Fig.~2. Measured photon spectrum in the reaction $^{181}Ta + ^{197}Au$ at 39
MeV/A is compared with the DCM calculations (the solid line). The calculated
spectrum is decomposed into fractions corresponding to the following elementary
processes : $pn \rightarrow pn\gamma $, $N\pi \rightarrow N\gamma $,
$pn \rightarrow d\gamma $. 
\end{small}}
\label{fig2}
\end{figure}
one may derive the limiting probability
distribution of the relative energy field :
\begin{equation}
\label{eq2}
P(\varepsilon ) = \frac{N}{\varepsilon} \exp \Bigg( - \frac{(\log{\varepsilon }
- \mu )^2}{2s} \Bigg)~~~ \ ,
\end{equation}
where $N$ is the normalization factor, $\sqrt{s}$~ is the width
of the Gaussian variable and $\mu \equiv \log(E_{CM}/2A) + s/2$~. The
production cross section of a particle '$x$' 
in the NN collisions is then :
\begin{equation}
\label{eq3}
{\sigma}^{(x)} = \int_{E_{th}^{(x)}}^{E_{tot}} d\varepsilon P(\varepsilon )
{\sigma}_{NN}^{(x)}(\varepsilon - E_{th}^{(x)}) ~~~\ ,
\end{equation}
where $E_{th}^{(x)}$ is the energy 
threshold in the NN collision and $E_{tot}$ is the
total energy available for the production of a particle 
in the HI collision. 
\begin{figure}[tbh]
\begin{center}
\leavevmode
\epsfxsize=14.3cm
\epsfverbosetrue
\epsfbox{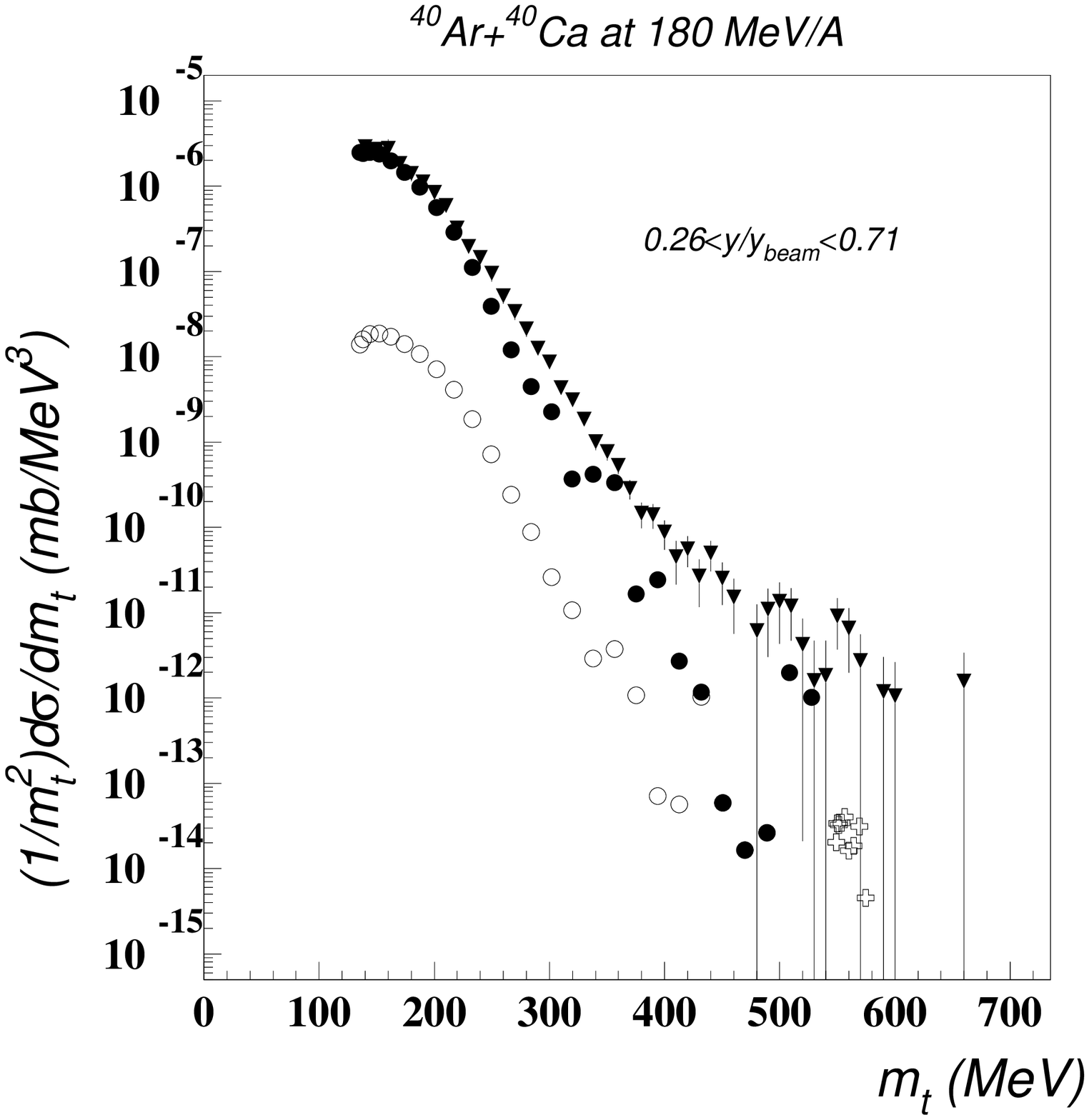}
%\epsffile{plo_fig3.eps}
\end{center}
{\begin{small} 
%\vspace*{-1cm}
Fig.~3. Experimental pion transverse mass distribution 
in the reaction $Ar + Ca$~ at 180
MeV/A is compared with those obtained in the DCM. For details see the
description in the text. 
\end{small}}
\label{fig3}
\end{figure}
Assuming that the time-duration of instabilities is
approximately independent of bombarding energy in a broad energy interval
and that the parameters of random, short-correlated medium depend weakly
both on energy and on the kind of produced particle, it was shown that all data
on total and differential cross sections of produced pions at $E_{CM} < 100
MeV/A$~ is remarkably well fitted by a distribution (\ref{eq2}) with  2 
parameters : $N$ and $s$ in the whole region\cite{guardamar}. 
Also the spectrum of deeply subthreshold $\eta$'s at
$E_{CM}/A = 180 MeV$ is reproduced.

The above stochastic multiplicative process is an attempt to
characterize the global features of relative
energy fluctuations in the off-equilibrium media
without entering into details of elementary processes involved in the
production of particle $x$. Alternatively, one can start from the
cascade picture of particle-particle collisions, including  
detailed informations about all elementary processes in the medium and,
in this way, one can 
try to reconstruct the particle production scenario in HI
reactions. Below, we shall use for this purpose the  Dubna Cascade Model (DCM)
\cite{dcm}~ which
has been recently extended by including various radiative capture processes
such as : $np \rightarrow d(\gamma , \pi , \eta \cdots )$~,  as well as the 
off-shell evolution of nucleons and $\Delta $'s\cite{gudima}~. 

\section{Subthreshold pion dynamics as a source of hard photons, pions and etas}
We shall omit discussion of ingredients concerning elementary processes
considered in DCM.  Interesting reader can
find further informations in Refs. \cite{gudima,taps}. 
In the calculations of the
photon spectrum, we consider the $pn$ bremsstrahlung process 
in the one-boson approximation\cite{schaeffer} and we omit the $pp$
bremsstrahlung. We include also the radiative capture process : $pn \rightarrow
d\gamma $~, which can be important in the description of high energy part
of the photon spectrum.  This process requires the
existence of bound state of final deuteron, i.e., the deuteron momentum must
be above the Mott momentum if the influence of the surrounding nucleons is
taken into account\cite{rozne,grupa}~. 
The process $pn \rightarrow d\gamma $ is most 
important at $E/A \simeq 150$MeV\cite{gudima,grupa}~. 
The decay processes : ${\pi }^0 \rightarrow \gamma \gamma $ 
and $\Delta \rightarrow N\gamma $~, which could give rise to
hard photons, have been included as well. 

It turns out that the most important
source of hard photons is the process : $\pi N \rightarrow N\gamma
$~\cite{taps}. Pions in the DCM are produced either directly : $NN \rightarrow
NN\pi $~ , or in two steps involving intermediate $\Delta $~
or off-shell nucleon\cite{bertsch} : $NN \rightarrow \Delta N$ or 
$NN \rightarrow N^{(off)}N$ and then
$\Delta \rightarrow N\pi $~, $\Delta N \rightarrow NN\pi $~
or $N^{(off)}N \rightarrow NN\pi $~. Formation of $\Delta $
is calculated so that the effective mass of the $\pi N$ system follows the
$\Delta $ mass distribution, its width being dependent on the pion momentum in
the $\pi N$ system. 
The yield of the primordial pions is subsequently modified
by the absorption and rescattering processes in the nuclear medium\cite{taps}.
The same two-step mechanism as well as
the process $\pi N^{(off)} \rightarrow N\pi (\eta \cdots )$~ have been 
considered as a source of hard pions and deeply subthreshold etas. 

Fig. \ref{fig1} shows the comparison of 
the experimental data for the ${\pi }^{+}$ production
in $p + Ar$ reaction\cite{jakobsson} in the broad range of energies with the 
results of DCM calculations. The agreement between theory and experiment 
becomes even more impressive if one notices that 
the experimental data contains a systematic error of about 40 \%
\cite{jakobsson}.

Fig. \ref{fig2} compares calculated and 
measured photon spectrum (full points) in the
reaction $^{181}Ta+^{197}Au$ at 39MeV/A \cite{taps} 
The dominant contributions : $pn \rightarrow pn\gamma $~, $pn \rightarrow
d\gamma $ and $N\pi \rightarrow N\gamma $~, are shown for the
comparison. The contribution from the decay ${\pi }^0 \rightarrow \gamma \gamma
$ is insignificant at this bombarding energy. The agreement
between DCM calculations of the photon spectrum and the experimental data is
excellent up to the highest photon energies.

Finally, Fig. \ref{fig3} presents measured\cite{taps1} 
(full triangles) and calculated
(full points) transverse mass spectrum of pions  for the reaction 
$Ar + Ca$~ at 180MeV/A. One may notice the deficit of hard pions 
($m_t > 0.4$GeV )~ in DCM calculations. The
measured spectrum decreases slower than exponentially, similarly as in
the log-normal behaviour. The open points in Fig. \ref{fig3} show the
contribution 
from processes involving off-shell nucleons. The crosses give the calculated
$\eta$-production cross section. The estimated 
meson ratio ${\sigma }_{\eta }/{\sigma }_{{\pi}^0}$ is $\leq 3 \cdot 10^{-6}$ in
the data as compared to $3 \cdot 10^{-7}$ in the DCM\cite{taps1}~.

\section{Imaging the final-state phase space in HI collisions by
HBT correlations}
Here we want to examine the status of HBT measurements\cite{hbt} 
as a tool to study the 
spatio-temporal extension of the emitting source at intermediate energy
HI collisions. The two-particle correlation function is defined as :
\begin{equation}
\label{eq4}
C_2^{ex}(p_1,p_2) \equiv \frac{P_2(p_1,p_2)}{P_1(p_1)P_1(p_2)} = 
1 + \frac{\mid \int
\exp (iqx) S(x,K) d^{4}x {\mid }^2}{[\int S(x_1,p_1)d^{4}x_1] \cdot
[\int S(x_2,p_2)d^{4}x_2]}~~~\ ,
\end{equation}
where $S$~ is the source term (the Wigner transform of the single particle
density matrix), $q \equiv p_2-p_1$ and $K \equiv (1/2) (E_1+E_2,~ {\bf
p_1}+{\bf p_2})$. $C_2^{ex}$ is positive definite even though $S(x,p)$~ may
take both positive and negative values. 
Eq. (\ref{eq4}) constitutes a formal basis for incorporating the Bose-Einstein
symmetrization in the cascade codes of HI reactions. However,
in the actual applications, one replaces this exact result
by\cite{pratt} :
\begin{equation}
\label{eq5}
C_2^{app}(p_1,p_2) = \frac{\mid \int \exp (iqx_1) S(x_1,p_1) d^4x_1 
\int \exp (iqx_2)
S(x_2,p_2) d^4x_2 \mid }{[ \int S(x_1,p_1) d^4x_1] \cdot [\int S(x_2,p_2)
d^4x_2]}
\end{equation}
which corresponds to weighting the pair at space-time points $(x_1,p_1)$~,
$(x_2,p_2)$~ with $1 + \cos [(p_1-p_2)(x_1-x_2)]$~, and is equivalent to the
assumption of smoothness of the source function.
This assumption may lead to certain pathological results (see, e.g., 
the discussion in Ref. \cite{zajc})~. 

The advantage of photons over other probes 
is that they escape from the interaction region unperturbed. Let us take 
the photon source in the form :
\begin{equation}
\label{eq6}
S(x,K) = N \exp (-E/T) \exp (- {\bf r}^2/R_0^2) \exp (- t^2/{\tau }^2) ~~~\ ,
\end{equation} 
which approximates well the source function extracted
from the DCM evolution \cite{gudima}~. 
The correlation function corresponding to (\ref{eq6}) is :
\begin{equation}
\label{eq7}
C_2^{ex}(Q, q_0) = 1 + \frac{2\Lambda }{z^2 K_2(z)}
{\exp \Bigg(- \frac{Q^2R_0^2}{2}\Bigg)}
\exp \Bigg(- \frac{q_0^2}{2}({\tau }^2 + R_0^2)\Bigg) ~~~ \ ,
\end{equation}
where $z \equiv Q/T$~, $Q^2 \equiv {\bf q}^2 - q_0^2$~ 
and $K_2(z)$~ is the McDonald function. The polarization factor\cite{neuhauser} 
:  $\Lambda = (1/4)(1+{\cos }^2{\Theta}_{12})$~, 
where ${\Theta}_{12}$ is the angle between photon
momenta, has been taken equal 1/2. Using (\ref{eq5}) one finds :
\begin{equation}
\label{eq8}
C_2^{app}(Q, q_0) = 1 + \Lambda \exp \Bigg( -\frac{Q^2R_0^2}{2}\Bigg)
\exp \Bigg(- \frac{q_0^2}{2}({\tau }^2 + R_0^2)\Bigg) ~~~ \ .
\end{equation}
Note the absence of the pre-exponential function, which 
is a measure of importance of the off-shell effects and
disappears if : $Q \rightarrow 0$~ or $T \rightarrow \infty$. 

\begin{figure}[tbh]
\begin{center}
\leavevmode
\epsfxsize=14.3cm
%\epsfysize=8.0cm
%\epsfverbosetrue
\epsfbox{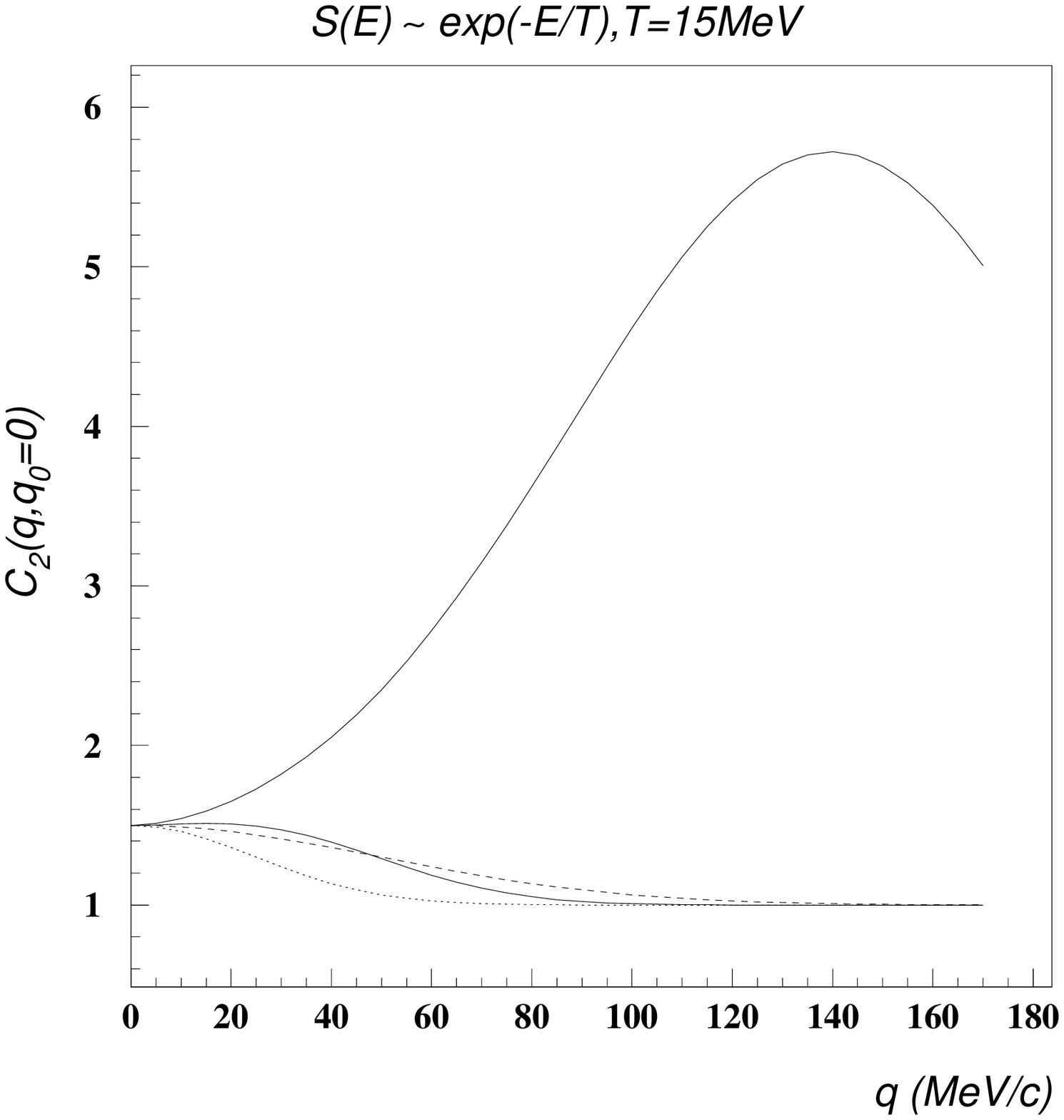}
%\epsffile{plo_fig4.eps}
\end{center}
%\vspace{-1cm}
{\begin{small} 
Fig.~4. Four different projections of the correlation function calculated
analytically for the source distribution (\ref{eq6}). The solid lines show
results obtained exactly from the definition (\ref{eq4}) for the two source
sizes : $R_0=4$ and 8fm, whereas the dashed and dotted curves have been 
obtained in the smoothness approximation (\ref{eq5}) for $R_0=4$ and 8fm
respectively.
\end{small}}
\label{fig4}
\end{figure}

Both exact (\ref{eq7}) and approximate (\ref{eq8}) 
correlation functions, are
plotted in Fig. \ref{fig4} for two source sizes : $R_0=4$fm and 8fm, and for the
'apparent temperature' $T=15$MeV,  which is a 
typical value in HI reactions at $E/A < 100$MeV. $C_2^{ex}$
for $R_0=4$fm shows a strong increase  
due to the fact that a typical photon wavelength in this low-$T$ regime 
is comparable to the source size. For both values of
$R_0$~ the approximate correlation function provides a poor approximation of
the exact one, i.e., the cascade calculations using the smoothness 
approximation (\ref{eq5}) are not reliable.
These calculations could be improved using the smearing
prescription\cite{heinz}~ where the 'classical' source is
convoluted with a minimum uncertainty Gaussian wave-packet : $g(x,p) \sim \exp
[- x^2/(2{\sigma}^2) + 2p^2{\sigma}^2)]$~. However, in this case
results both for $C_2$~ and for the energy spectra depend on the
value of the smearing parameter ${\sigma}^2$~. 

In the experimental analysis,
the source size could be extracted more reliably 
using Eq. (\ref{eq7}) in the fitting procedure. 
One should however notice that for small
source sizes, the rise of the correlation function can be easily 
confused with a Gaussian peak at $Q \sim {m_{\pi}}$~
from the decay : ${\pi}^{0} \rightarrow \gamma \gamma$~. 
In this unfavorable case, the experimental extraction of $R_0$~
may be very difficult.

\section{Conclusions}
We have shown that cascade dynamics with the schematic account of
mean-field dynamics, is successful in reproducing spectra of extremely high
energy photons and hard pions produced 
in the deeply subthreshold regime of HI 
reactions. The significant deviations
between the DCM calculations and the data begins to be visible only at 
extremely large $m_t$'s. More data in this domain, which 
may provide a stringent test of the time-evolution of energy fluctuations in
the off-equilibrium nuclear medium, are badly needed.

In principle, the same cascade scenario could be tested also in 
HBT measurements. However, the rapid 
shape variations of the $\gamma \gamma $- correlation function 
with the source radius and the 
importance of off-shellness ($Q \neq 0$) in the  
intermediate energy HI reactions may pose serious problems in extracting
the quantitative information. HBT measurements
using pions in this domain are also not free from these defects but the
corrections in this case are somewhat easier to introduce 
both in the calculations and in the experimental analysis\cite{gudima}~.

\vfill
\newpage

\bibliographystyle{unsrt}

\end{document}